# Survey of Security and Privacy Issues of Internet of Things

*Tuhin Borgohain
Department of Instrumentation Engineering, Assam Engineering College, Guwahati-13
Email: borgohain.tuhin@gmail.com
**Uday Kumar**
Delivery Manager, Tech Mahindra Limited, India
Email: udaykumar@techmahindra.com
**Sugata Sanyal**
Corporate Technology Office, Tata Consultancy Services, Mumbai, India
Email: sugata.sanyal@tcs.com
*Corresponding author

--------------------------------------------------------------------ABSTRACT-------------------------------------------------------------
This paper is a general survey of all the security issues existing in the Internet of Things (IoT) along with an analysis of the privacy issues that an end-user may face as a consequence of the spread of IoT. The majority of the survey is focused on the security loopholes arising out of the information exchange technologies used in Internet of Things. No countermeasure to the security drawbacks has been analyzed in the paper.

## 1. INTRODUCTION

Building upon the concept of Device to Device (D2D) communication technology of Bill Joy [1], Internet of Things (IoT) embodies the concept of free flow of information amongst the various embedded computing devices using the internet as the mode of intercommunication. The term "Internet of Things" was first proposed by Kevin Ashton in the year 1982 [2]. With the aim of providing advanced mode of communication between the various systems and devices as well as facilitating the interaction of humans with the virtual environment, IoT finds its application in almost any field. But as with all things using the internet infrastructure for information exchange, IoT to is susceptible to various security issues and has some major privacy concerns for the end users. As such IoT, even with all its advanced capabilities in the information exchange area, is a flawed concept from the security viewpoint and proper steps has to be taken in the initial phase itself before going for further development of IoT for an effective and widely accepted adoption.

## 2. OVERVIEW

In section 3 of this paper we discuss the various communication technologies using the Internet infrastructure for the exchange of information. In section 4, we do a survey of all the security issues plaguing the Internet of Things as well as the pervading privacy issues faced by the end users of technologies utilizing the advanced information sharing architecture of IoT. In section 5, we conclude our paper with a proposal for the necessary steps to be taken for addressing all the security issues of IoT.

## 3. CONNECTIVITY TECHNOLOGIES AND INTERACTION AMONGST VARIOUS INTERNET OF THINGS (IoT) DEVICES

The automatic exchange of information between two systems or two devices without any manual input is the main objective of the Internet of Things. This automated information exchange between two devices takes place through some specific communication technologies, which are described below.

### 3.1 Wireless Sensor Networks (WSN)

As described in [3], WSN are compositions of independent nodes whose wireless communication takes place over limited frequency and bandwidth. The communicating nodes of a typical wireless sensor network consist of the following parts:
 i. Sensor
 ii. Microcontroller
 iii. Memory
 iv. Radio Transceiver
 v. Battery

Due to the limited communication range of each sensor node of a WSN, multi-hop relay of information take place between the source and the base station. The required data is collected by the wireless sensors through collaboration amongst the various nodes, which is then sent to the sink node for directed routing towards the base station. The communication network formed dynamically by the use of wireless radio transceivers facilitates data transmission between nodes. Multi-hop transmission of data demands different nodes to take diverse traffic loads [2].

*3.2 Radio Frequency Identification (RFID)*

In context to the Internet of Things (IoT), RFID technology is mainly used in information tags interacting with each other automatically. RFID tags use radio frequency waves for interacting and exchanging information between one another with no requirement for alignment in the same line of sight or physical contact. It uses the wireless technology of Automatic Identification and Data Capture (AIDC) [23]. A RFID is made up of the following two components [2]:

*3.2.1 RFID tags (Transponders)*

In a RFID tag, an antenna is embedded in a microchip. The RFID tag also consists of memory units, which houses a unique identifier known as Electronic Product Code (EPC). The function of the EPC in each tag is to provide a universal numerical data by which a particular tag is recognized universally.

As per the classification in [2], the types of RFID tags are:

i. Active tag: This type of tag houses a battery internally, which facilitates the interaction of its unique EPC with its surrounding EPCs remotely from a limited distance.

ii. Passive tag: In this type of tag, the information relay of its EPC occurs only by its activation by a transceiver from a pre-defined range of the tag. The lack of an internal battery in the passive tags is substituted by its utilization of the electromagnetic signal emitted by a tag reader through inductive coupling as a source of energy. (For details about the utilization of external sources of energy in a passive tag, readers can refer to [4]).

A RFID tag operates in conjunction with a tag reader, the EPC of the former being the identifying signature of a particular tag under the scan of the latter.

*3.2.2 RFID readers (Transceivers)*

The RFID reader functions as the identification detector of each tag by its interaction with the EPC of the tag under its scan.

More information on the working technologies behind RFID can be found in [6].

## 4. SECURITY ISSUES AND PRIVACY CONCERNS

Despite the immense potential of IoT in the various spheres, the whole communication infrastructure of the IoT is flawed from the security standpoint and is susceptible to loss of privacy for the end users. Some of the most prominent security issues plaguing the entire developing IoT system arise out of the security issues present in the technologies used in IoT for information relay from one device to another. As such some of the prominent security issues stemming out from the communication technology are the following:

*4.1 Security issues in the wireless sensor networks (WSNs):*

The hierarchical relationship of the various security issues plaguing the wireless sensor network is shown in Figure 1. The oppressive operations that can be performed in a wireless sensor network can be categorized under three categories [7]:

    i. Attacks on secrecy and authentication

    ii. Silent attacks on service integrity

    iii. Attacks on network availability: The denial of service (DoS) ([16], [17]) attack falls under this category. This prevention of accessibility of information to legitimate users by unknown third party intruders can take place on different layers of a network [8],[14],[15]:

*4.2 DoS attack on the physical layer*:

The physical layer of a wireless sensor network carries out the function of selection and generation of carrier frequency, modulation and demodulation, encryption and decryption, transmission and reception of data [19]. This layer of the wireless sensor network is attacked mainly through

    i. Jamming: In this type of DoS attack occupies the communication channel between the nodes thus preventing them from communicating with each other.

    ii. Node tampering: Physical tampering of the node to extract sensitive information is known as node tampering.

*4.3 DoS attack on the link layer*:

The link layer of WSN multiplexes the various data streams, provides detection of data frame, MAC and error control. Moreover the link layer ensures point-point or point-multipoint reliability [20]. The DoS attacks taking place in this layer are:

    i. Collision: This type of DoS attack can be initiated when two nodes simultaneously transmit packets of data on the same frequency channel. The collision of data packets results in small changes in the packet results in identification of the packet as a mismatch at the receiving end. This leads to discard of the affected data packet for re-transmission [22].

    ii. Unfairness: As described in [22], unfairness is a repeated collision based attack. It can also be referred to as exhaustion based attacks.

    iii. Battery Exhaustion: This type of DoS attack causes unusually high traffic in a channel making its accessibility very limited to the nodes. Such a disruption in the channel is caused by a large number of requests (Request To Send) and transmissions over the channel.

*4.4 DoS attack on the network layer*:

The main function of the network layer of WSN is routing. The specific DoS attacks taking place in this layer are:

    i. Spoofing, replaying and misdirection of traffic.

    ii. Hello flood attack: This attack causes high traffic in channels by congesting the channel with an unusually high number of useless messages. Here a single malicious node sends a useless message which is then replayed by the attacker to create a high traffic.

    iii. Homing: In case of homing attack, a search is made in the traffic for cluster heads and key managers which have the capability to shut down the entire network.

    iv. Selective forwarding: As the name suggests, in selective forwarding, a compromised node only sends a selected few nodes instead of all the nodes. This selection of the nodes is done on the basis of the requirement of the attacker to achieve his malicious objective and thus such nodes does not forward packets of data.

    v. Sybil: In a Sybil attack, the attacker replicates a single node and presents it with multiple identities to the other nodes.

    vi. Wormhole: This DoS attack causes relocation of bits of data from its original position in the network. This relocation of data packet is carried out through tunnelling of bits of data over a link of low latency.

vii. Acknowledgement flooding: Acknowledgements are required at times in sensor networks when routing algorithms are used. In this DoS attack, a malicious node spoofs the Acknowledgements providing false information to the destined neighboring nodes

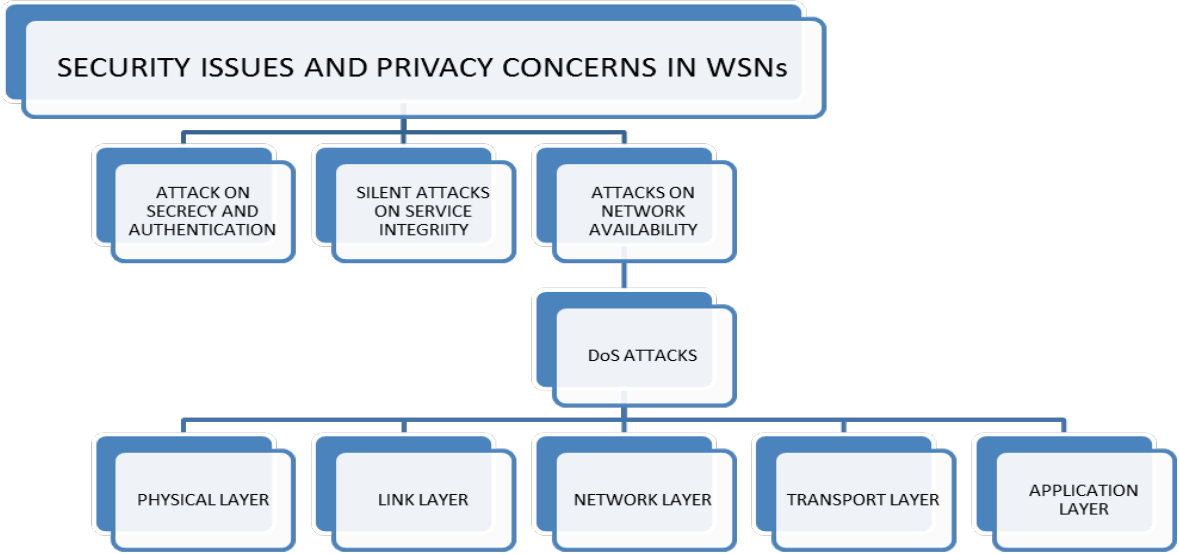

**Figure 1 - Hierarchical diagram of security issues in Wireless Sensor Network**

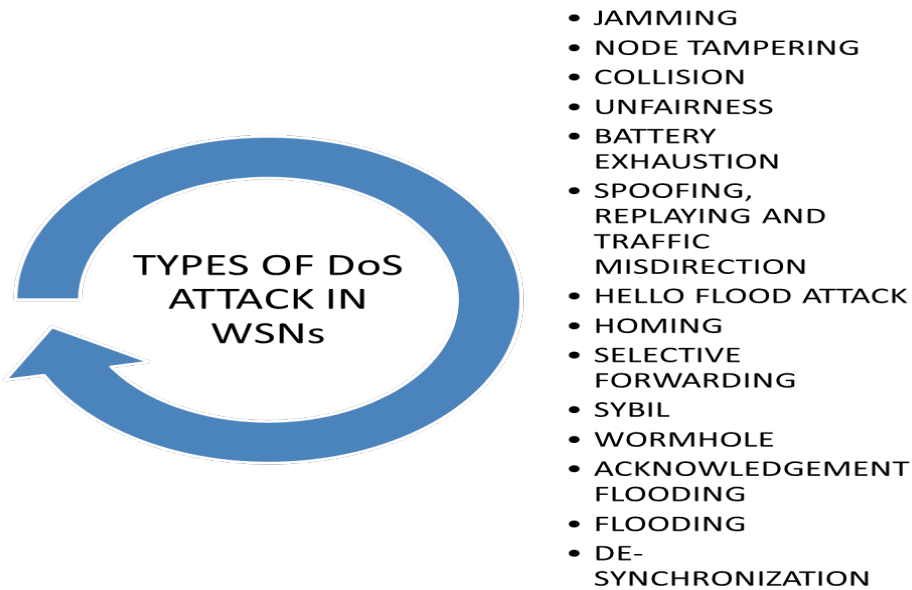

**Figure 2 - Types of Denial of Attack in Wireless Sensor Network**

*4.4 DoS attack on the transport layer*:

This layer of the WSN architecture provides reliability of data transmission and avoids congestion resulting from high traffic in the routers. The DoS attacks in this layer are:

  i. Flooding: It refers to deliberate congestion of communication channels through relay of unnecessary messages and high traffic.

  ii. De-synchronization: In de-synchronization attack, fake messages are created at one or both endpoints requesting retransmissions for correction of non-existent error. This results in loss of energy in one or both the end-points in carrying out the spoofed instructions.

*4.5 DoS attack on the application layer*:

The application layer of WSN carries out the responsibility of traffic management. It also acts as the provider of software for different applications which carries out the translation of data into a comprehensible form or helps in collection of information by sending queries [20]. In this layer, a path-based DoS attack is initiated by stimulating the sensor nodes to create a huge traffic in the route towards the base station [21], [22].

Figure 2 shows all the above mentioned DoS attacks in the different layers of a wireless sensor network.

Some additional DoS attacks are as follows [7], [14], [15], [36]:
  i. Neglect and Greed Attack
  ii. Interrogation
  iii. Black Holes
  iv. Node Subversion
  v. Node malfunction
  vi. Node Outage
  vii. Passive Information Gathering
  viii. False Node
  ix. Message Corruption

Some of the other security and privacy issues in a WSN are [7], [9], [10]:
  i. Data Confidentiality
  ii. Data Integrity
  iii. Data Authentication
  iv. Data Freshness
  v. Availability
  vi. Self-Organization
  vii. Time Synchronization
  viii. Secure Localization
  ix. Flexibility
  x. Robustness and Survivability

According to [26], the threats looming over WSN can further be classified as follows:
  i. External versus internal attacks
  ii. Passive versus active attacks
  iii. Mote-class versus laptop-class attacks

According to [12], the attacks on WSN can be classified as:
  i. Interruption
  ii. Interception
  iii. Modification
  iv. Fabrication

The attacks on WSN can further be classified as:
  i. Host-based attacks
  ii. Network-based attacks

*4.6 Security issues in RFID technology*

In context to IoT, RFID technology is mainly used as RFID tags for automated exchange of information without any manual involvement. But the RFID tags are prone to various attacks from outside due to the flawed security status of the RFID technology. The four most common types of attacks and security issues of RFID tags ([25], [35]) are shown in Figure 3 which are as follows:

i. Unauthorized tag disabling (Attack on authenticity): The DoS attacks in the RFID technology leads to incapacitation of the RFID tags temporarily or permanently. Such attacks render a RFID tag to malfunction and misbehave under the scan of a tag reader, its EPC giving misinformation against the unique numerical combination identity assigned to it. These DoS attacks can be done remotely, allowing the attacker to manipulate the tag behavior from a distance.

ii. Unauthorized tag cloning (Attack on integrity): The capturing of the identification information (like its EPC) esp. through the manipulation of the tags by rogue readers falls under this category. Once the identification information of a tag is compromised, replication of the tag (cloning) is made possible which can be used to bypass counterfeit security measures as well as introducing new vulnerabilities in any industry using RFID tags automatic verification steps [35].

iii. Unauthorized tag tracking (Attack on confidentiality): A tag can be traced through rogue readers, which may result in giving up of sensitive information like a person's address. Thus from a consumer's viewpoint, buying a product having an RFID tag guarantees them no confidentiality regarding the purchase of their chase and in fact endangers their privacy.

iv. Replay attacks (Attack on availability): In this type of impersonation attacks the attacker uses a tag's response to a rogue reader's challenge to impersonate the tag [25]. In replay attacks, the communicating signal between the reader and the tag is intercepted, recorded and replayed upon the receipt of any query from the reader at a later time, thus faking the availability of the tag.

Besides this category, some prominent security vulnerabilities of RFID technologies are [35]:
  i. Reverse Engineering
  ii. Power Analysis
  iii. Eavesdropping
  iv. Man-in-the-middle attack
  v. Denial of Service (DoS)
  vi. Spoofing
  vii. Viruses
  viii. Tracking
  ix. Killing Tag Approach

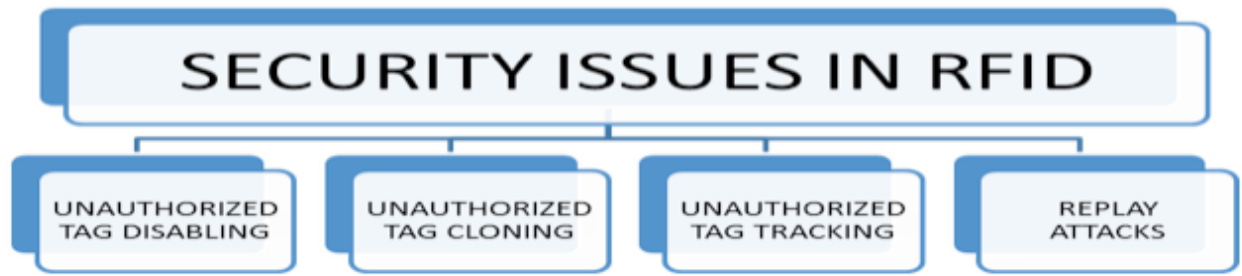

Figure 3 – Security Issues in RFID

*4.7 Security issues in health-related technologies built upon the concept of IoT:*

Advances and convergence of engineering with biology has paved the way for wearable health monitoring devices which can constantly stream and share the information from the sensor of the health monitor with other devices and social network over the internet (The implementation of social connectivity with the sensor data can be found in [28], [30] and [31]). The implementation of automatic collection of data by the sensors and uploading it to the various social networks through a web server introduces some high vulnerability in the whole data transmission process from the monitor to the Internet. On the basis of its target device (FITBIT), the authors of ([27], [32]) have recognized the following as the main security vulnerability in such health monitoring devices working in synchronization with the Internet:

   i. Clear text login information: During login to the account linked with the health monitoring device, the authenticated password of the user is registered in the web server in clear text which is then recorded in log files. This gives way to loss of secured login by making the password available easily through the log files.

   ii. Clear text HTTP data processing: The sensor data is sent to the web servers as plain HTTP instructions with no additional security or encryption. Such unprotected HTTP instructions can be easily intercepted for gaining access to various functions of a user account linked to the health-monitoring device.

From the above mentioned vulnerabilities it is clear that the security measures implemented in the health-related technologies which are socially connected over the internet lack the proper measures to address all the privacy concerns of the end users and puts the users at risk of exposing valuable information about their health to unknown personnel with malicious intents.

Based on the above-mentioned security flaws, many other security and privacy issues present themselves in the field of Internet of Things. A few of them are:

   i. Theft of sensitive information like bank password
   ii. Easy accessibility to personal details likes contact address, contact number etc.
   iii. It may lead to open access to confidential information like financial status of an institution
   iv. An attack on any one device may compromise the integrity of all the other connected devices. Thus the interconnectivity has a huge drawback as a single security failure can disrupt an entire network of devices.
   v. The reliance on the Internet makes the entire IoT architecture susceptible to virus attack, worm attack and most of the other security drawbacks that comes with any Internet connected computing device etc.

## 5. CONCLUSION

In this paper we have surveyed all the security flaws existing in the Internet of Things that may prove to be very detrimental in the development and implementation of IoT in the different fields. So adoption of sound security measures ([18], [24], [29], [34]) countering the above detailed security flaw as well as implementation of various intrusion detection systems ([11], [33]), cryptographic and stenographic security measures ([5]) in the information exchange process and using of efficient methods for communication ([13]) will result in a more secure and robust IoT infrastructure. In conclusion, we would like to suggest that more effort on development of secured measures for the existing IoT infrastructure before going for further development of new implementation methods of IoT in daily life would prove to be a more fruitful and systematic method.

**Biographies and Photographs**

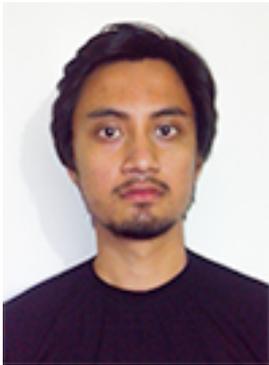

Tuhin Borgohain is a 3rd Year student of Assam Engineering College, Guwahati. He is presently pursuing his Bachelor of Engineering degree in the department of Instrumentation Engineering.

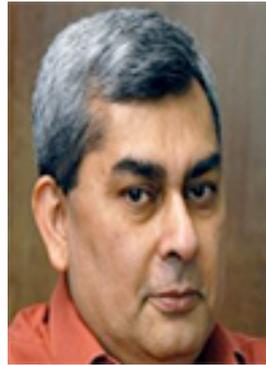

Sugata Sanyal is presently acting as a Research Advisor to the Corporate Technology Office, Tata Consultancy Services, India. He was with the Tata Institute of Fundamental Research till July, 2012. Prof. Sanyal is a: Distinguished Scientific Consultant to the International Research Group: Study of Intelligence of Biological and Artificial Complex System, Bucharest, Romania; Member, "Brain Trust," an advisory group to faculty members at the School of Computing and Informatics, University of Louisiana at Lafayette's Ray P. Authement College of Sciences, USA; an honorary professor in IIT Guwahati and Member, Senate, Indian Institute of Guwahati, India. Prof. Sanyal has published many research papers in international journals and in International Conferences worldwide: topics ranging from network security to intrusion detection system and more.

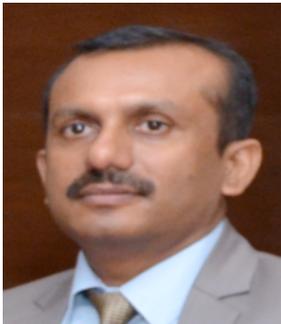

Uday Kumar is working as Delivery Manager at Tech Mahindra Ltd, India. He has 17 years of experience in engineering large complex software system for customers like Citibank, FIFA, Apple Computers and AT&T. He has developed products in BI, performance testing, compilers. And have successfully led projects in finance, content management and ecommerce domain. He has participated in many campus connect program and conducted workshop on software security, skills improvement for industrial strength programming, evangelizing tools and methodology for secure and high end programming.